\documentclass[11pt]{article}
\usepackage{epsfig,sint,macros}
\begin{document}

\begin{titlepage}
\begin{flushright}
  COLO-HEP-480 \\
  HU-EP-02/08
\end{flushright}

\vskip 1 cm
\begin{center}
  {\Large\bf Simulation of dynamical fermions with smeared links }
\end{center}
\vskip 1 cm
\begin{center}
{\large Anna Hasenfratz $^{a~\dagger}$ and Francesco Knechtli $^{b~\ddagger}$}
\vskip 0.8cm
$^{a}$ Physics Department, 
                     University of Colorado, Boulder, CO 80309 USA
\vskip 0.4cm
$^{b}$ Institut f{\"u}r Physik, Humboldt-Universit{\"a}t zu Berlin,\\
                     Invalidenstr. 110, 10115 Berlin, Germany
\end{center}
\vskip 2.5ex
{\bf Abstract}
\vskip 0.7ex
Smearing the gauge links of dynamical configurations removes small scale
unphysical vacuum fluctuations und thus improves the chiral properties of
lattice fermions. We present a new algorithm for the simulation of dynamical
fermions coupled via smeared links based on the standard pure gauge overrelaxation
and heatbath updatings. Smeared links play a fundamental role in making this
algorithm effective. At fixed lattice spacing the computational cost of
the algorithm has an extra volume factor due to the finite volume of the lattice
region which can be updated. As the continuum limit is approached the physical
volume of the updated region remains constant.
We simulated four flavors of staggered fermions coupled via hypercubic (HYP)
smeared links. The simulation cost of the new algorithm on $10$fm$^4$ volumes is a
factor 2--8 larger
than with the standard Hybrid Monte Carlo but the improved properties of the HYP
action allow to gain a factor 2 in the lattice spacing. The new algorithm could be
applicable to simulations of more complicated chiral fermionic actions,
like overlap or perfect actions.
\vfill

\begin{flushleft}
  COLO-HEP-480 \\
  HU-EP-02/08 \\
  March 2002
\end{flushleft} 
$^{\dagger}$ {\small e-mail: anna@eotvos.colorado.edu} \\
$^{\ddagger}$ {\small e-mail: knechtli@physik.hu-berlin.de}
\eject

\vfill

\eject

\end{titlepage}


\section{Introduction \label{s_intro}}

In this article we propose an algorithm for dynamical simulations of
smeared link fermions.
We consider a system described by the action
\begin{equation}\label{action}
S = S_g(U)-{\rm tr}\ln\left[Q^{\dagger}(V)Q(V)\right] \,,
\end{equation}
where $S_g(U)$ is the pure gauge action and $Q(V)$ is the fermionic matrix.
The gauge connections between the fermions are smeared links $V$
which are constructed deterministically 
from the dynamical thin links $U$. Since each
smeared link is a local combination of a finite number of thin links,
the system where the
fermions couple via smeared links is in the same universality 
class as the system with thin links.
In the action we keep the smearing of the links unchanged as the 
continuum limit is approached.

Smearing of the gauge links in the action and operators of a theory is part of
many improvement programs. 
It has been demonstrated that smeared links improve flavor symmetry with staggered
fermions \cite{Orginos:1998ue,Orginos:1999cr,Knechtli:2000ku} and
chiral symmetry with Wilson-type clover fermions \cite{DeGrand:1998mn}.
They are useful in the construction of an overlap Dirac operator
\cite{Bietenholz:2000iy,DeGrand:2000tf,DeGrand:2000gq} and arise naturally
in the framework of perfect actions
\cite{DeGrand:1998pr,Niedermayer:2000yx,Hasenfratz:2001hr}.
As an example for improved operators we mention the classically
perfect Polyakov loop in pure gauge theory \cite{DeGrand:1995ji}. In order to
eliminate lattice artifacts the static source has to become an extended object. 
Wilson loops constructed from smeared links were found to improve the statistical
accuracy of potential measurements in \cite{Hasenfratz:2001tw,Gattringer:2001jf}
also.

The problem with eq. (\ref{action}) is that the standard simulation
algorithms for dynamical fermions (Hybrid Monte Carlo (HMC) or R algorithms)
involve the computation of the gauge force $\partial Q(V)/\partial U$, which is
either very complicated or likely impossible if the smeared links are
projected onto SU(3). Present large scale dynamical simulations with standard
algorithms \cite{Karsch:2000ps,Bernard:2001av} use one level of
non--projected smearing.
Quenched studies indicate that it would be desirable to simulate dynamical
fermions with smoother links, either obtained by a more complex and effective
smearing procedure like the hypercubic (HYP) blocking \cite{Hasenfratz:2001hp}
or by iterative smearing with projection onto SU(3) after each smearing step.

\section{The HYP action \label{s_hyp}}

The definition of HYP links and their properties are discussed in
\cite{Hasenfratz:2001hp,Hasenfratz:2001tw}.
In the natural formulation of four flavors of staggered quarks, the continuum
SU(4) flavor symmetry is broken to U(1) at finite lattice spacing. The pion
spectrum has only one true Goldstone pion $\pi_{\rm G}$, the other 14 pions remain
massive when the quark mass approaches zero. The mass splitting between a
non--Goldstone pion $\pi$ and $\pi_{\rm G}$ can be parametrized by the quantity
\cite{Orginos:1998ue}
\begin{equation}\label{delta2}
\delta_2 =
\frac{m_{\pi}^2-m_{\rm G}^2}{m_{\rho}^2-m_{\rm G}^2}
\end{equation}
evaluated at a fixed $m_{\rm G}/m_{\rho}$ ratio.
With a flavor symmetric action $\delta_2=0$ for all the pions.

We computed $\delta_2$ at $m_{\rm G}/m_{\rho}=0.55$ 
for the two lightest non-Goldstone pions,
$\pi_{i,5}$ with flavor structure $\gamma_i\gamma_5$ and $\pi_{i,j}$
with flavor structure $\gamma_i\gamma_j$, on a set of quenched
$8^3\times24$ lattices generated with Wilson pure gauge action at $\beta=5.7$
($a=0.17\,{\rm fm}$),
using three different valence quark actions: standard staggered action with thin
or HYP smeared links and improved Asqtad action of the MILC collaboration
\cite{Orginos:1999cr}. The results are shown in table \ref{tdelta2}. 
Flavor symmetry violations with the HYP action are reduced by close to an order of
magnitude with respect to the thin link action and by a factor of two with
respect to the Asqtad action.
\begin{table}[h]
\vspace{0.0cm}
 \caption{Flavor symmetry breaking for different valence quark actions at
 lattice spacing $a=0.17\,{\rm fm}$. The parameter $\delta_2$ is defined in
 eq. (\ref{delta2}) and is evaluated at $m_{\rm G}/m_{\rho}=0.55$. \label{tdelta2}}
\vspace{0.2cm}
 \begin{tabular}{|c|c|c|c|} \hline
  Action/$\delta_2$ & Thin & Asqtad & HYP
  \\ \hline\hline
  $\pi_{i,5}$ & 0.594(25) & 0.191(22) & 0.086(14) \\ \hline
  $\pi_{i,j}$ & 0.72(6) & 0.32(4) & 0.150(24) \\ \hline
 \end{tabular}
\vspace{0.0cm}
\end{table}
\begin{figure}[tb]
\hspace{0cm}
\vspace{-1.0cm}
\centerline{\psfig{file=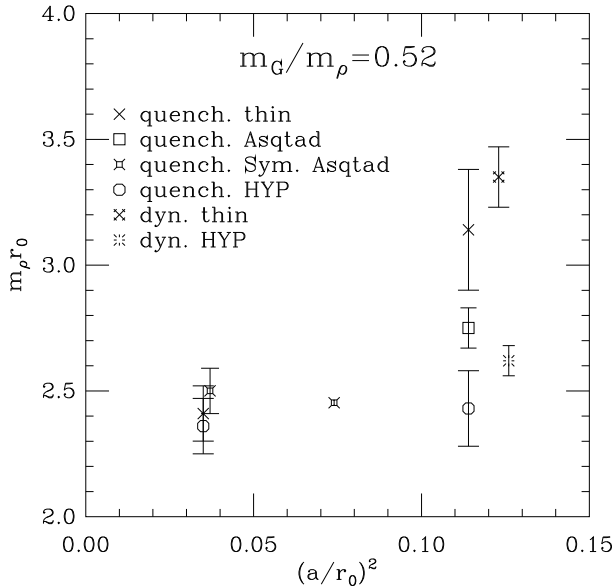,width=8cm}}
\vspace{0.2cm}
\caption{Scaling behavior of different staggered actions. All data are extrapolated to
$m_G/m_{\rho}=0.52$ except the Asqtad action with Symanzik improved gauge action
which is at $m_G/m_{\rho}=0.538(2)$ and the dynamical thin link action which is at
$m_G/m_{\rho}=0.56(2)$.
\label{f_scaling}}
\end{figure}

The HYP blocking was designed to improve flavor symmetry
by reducing the fluctuations of the gauge links that couple to the different
flavor and Dirac components  of staggered fermions inside
a hypercube. The parameters of the HYP transformation have been optimized
non--perturbatively
to minimize the short distance fluctuations of the smeared gauge field
\cite{Hasenfratz:2001hp}. A first non--trivial check of the HYP construction comes from perturbation
theory: The cancellation of the flavor symmetry violating terms at tree level 
in the HYP staggered action gives consistent results for the HYP parameters
\cite{annapert}. We looked further at the scaling properties of the HYP action.
Figure \ref{f_scaling} shows the rho mass as a function of the lattice spacing
squared, all in units of $r_0$ \cite{Sommer:1994ce}. In order to compare a
result from a dynamical HYP 
simulation 
at gauge coupling $\beta=5.1$ and bare quark mass $am=0.04$, we chose 
to extrapolate the quenched data to
$m_{\rm G}/m_{\rho}=0.52$, a slightly different value than the one used 
in table \ref{tdelta2}. The quenched results correspond to $\beta=5.7$ 
($r_0/a=2.96(2)$) and $\beta=6.0$ ($r_0/a=5.36(4)$) with the Wilson gauge action
and to $\beta=8.0$ ($r_0/a=3.687(13)$) and $\beta=8.4$ ($r_0/a=5.191(16)$)
with the Symanzik improved gauge action and Asqtad quark action \cite{dougasqtad}.
At $\beta=6.0$ we used for the thin link action data from ref. \cite{Gupta:1991mr}.
The quenched points show clearly the good scaling properties of the HYP action.
The available dynamical data are in agreement with the quenched ones.

\section{The New Algorithm \label{s_algo}}
 
We propose a two--step updating to simulate the system described by an action $S$ as given in eq. (\ref{action})
\begin{itemize}
\item[1)] update a set of thin links $\{U\}\rightarrow\{U^{\prime}\}$ 
in a way that 
satisfies detailed balance for $S_g(U)$
\item[2)] accept/reject the proposed change with acceptance probability
\begin{equation}\label{accrej}
P_{\rm acc} = \min\left\{1,
  \frac{\det\left[Q^{\dagger}(V^{\prime})Q(V^{\prime})\right]}
       {\det\left[Q^{\dagger}(V)Q(V)\right]}\right\} \,,
\end{equation}
where $V^{\prime}$ are the smeared links constructed from the updated thin links.
\end{itemize}
Since the smeared links $V$ are constructed deterministically form the thin links
$U$, they are not dynamical variables and an ergodic update of the $U$ links
will correctly simulate the system.
The proof that the above two--step updating
satisfies detailed balance for the action $S$
follows closely the proof given in \cite{Hasenbusch:1998yb} for a similar
two--step decomposition of the action.

In the practical implementation of the algorithm,
our choice for $S_g$ is the Wilson plaquette action and the
updating in step 1) is performed either with microcanonical overrelaxation 
\cite{Adler:1981sn,Petronzio:1990vx} or
with Cabibbo-Marinari heatbath \cite{Cabibbo:1982zn}.
We choose the subset
of thin links to be updated differently for the overrelaxation and
for the heatbath steps. For the overrelaxation we reflect all the
links within some finite block of the lattice. The location of the
block is chosen randomly but its dimensions are fixed. We choose with
probability 1/2 a given sequence of reflections within this block
or with equal probability the reversed sequence. The sequence has
to be reversed with respect to the direction and location of the thin
links and with respect to the index of the SU(2) subgroup used in
the reflection step. For the heatbath we choose the subset of thin
links randomly, i.e. we choose a random direction, a random parity,
a set of random sites and a random sequence of SU(2) subgroups. The
probability of generating a given sequence of thin link updates or
the reversed sequence is then the same, both for overrelaxation and
heatbath. From this it follows that detailed balance with respect
to the pure gauge action $S_g$ is satisfied for both updatings.
We chose to update a contiguous block of links with overrelaxation
in order to propagate a change through (a portion of) the lattice.
We observed that a random sequence of overrelaxed reflections is less
efficient, especially for changing the topological charge of the configurations.
This can be understood by thinking of instantons as extended objects.
One has to change an entire region of the lattice to destroy or to
create them.

In step 2) we use a stochastic estimator to evaluate the ratio of fermionic
determinants
\begin{eqnarray}
P_{\rm acc}^{\prime} & = & \min\left\{1,\exp{\Delta S}\right\} \label{pacc1}
\\
\Delta S & = & 
\xi^{\dagger}[Q^{\dagger}(V^{\prime})Q(V^{\prime})-Q^{\dagger}(V)Q(V)]\xi \,,
\label{pacc2}
\end{eqnarray}
where the vector $\xi$ is generated according to the probability distribution
\begin{equation}\label{pacc3}
P(\xi) \propto \exp\{-\xi^{\dagger}Q^{\dagger}(V^{\prime})Q(V^{\prime})\xi\} \,.
\end{equation}
The key feature is that smeared links constrain the {\em statistical fluctuations}
of the stochastic estimator and make the algorithm effective, as we will 
show
in the next section for the case of HYP staggered fermions.

Yet another interesting question is about the {\em actual value} of the ratio
of fermionic determinants in eq. (\ref{accrej}). This value is the
{\em natural limit} on the performance of the algorithm, something like the Carnot
efficiency of an heat engine. No matter how the accept/reject step 2) is performed,
the detailed balance condition implies that the maximal acceptance is given by the
Metropolis acceptance eq. (\ref{accrej})
\footnote{We thank U. Wolff for emphasizing this point.}.
We use a stochastic estimator to compute the acceptance and this reduces
in principle the efficiency of the algorithm. The actual value of the ratio
of fermionic determinants depends on the change in the gauge configuration.
A larger number of changed thin links $U$ leads to a smaller value
of the determinant ratio.

\section{Four flavors of HYP staggered fermions \label{s_4f}}

In the following we describe the application of the new algorithm to the case of
$n_f=4$ flavors of staggered fermions coupled via hypercubic (HYP)
smeared links.
The matrix $Q$ in eq. (\ref{action}) is given by
\begin{eqnarray}
Q_{i,j}(V) & = &
2(am)\delta_{i,j}+D_{i,j}(V) \,, \label{matrixnf4} \\
D_{i,j}(V) & = & 
\sum_{\mu}\eta_{i,\mu}(V_{i,\mu}\delta_{i,j-\hat{\mu}}-
 V^{\dagger}_{i-\hat{\mu},\mu}\delta_{i,j+\hat{\mu}}) \,. \nonumber
\end{eqnarray}
If we take the matrix $Q^{\dagger}Q$ to be defined on the even sites of the
lattice only \cite{Martin:1985yn},
eq. (\ref{action}) describes four flavors of staggered fermions coupled via HYP
smeared links $V$ ($\eta_{i,\mu}$ are the staggered phases).
To further reduce the fluctuations of the stochastic
estimator in the accept/reject step eq. (\ref{pacc1}), we remove from
the fermion matrix its most ultraviolet part by decomposing it as
\begin{eqnarray}
 Q(V) & = & Q_r(V)A(V) \,, \label{split} \\
 A(V) & = & \exp[\alpha_4D^4(V)+\alpha_2D^2(V)] \,. \label{amat}
\end{eqnarray}
This way we achieve that an effective action
\begin{equation}\label{seffD4}
S_{\rm eff} =
-2\alpha_4\rm{Re}\,{\rm tr}D^4
-2\alpha_2\rm{Re}\,{\rm tr}D^2
\end{equation}
is removed from the fermionic determinant. The acceptance probability 
in eq. (\ref{pacc1}) becomes
\begin{eqnarray}
\tilde{P}_{\rm acc} & = & \min\left\{1,
\exp[S_{\rm eff}(V)-S_{\rm eff}(V^{\prime})]\,\exp\Delta S_r\right\} 
\label{tildepacc} \\
\Delta S_r & = & 
\xi_r^{\dagger}[Q_r^{\dagger}(V^{\prime})Q_r(V^{\prime})-
                Q_r^{\dagger}(V)Q_r(V)]\xi_r \,, \label{deltaS_r}
\end{eqnarray}
where the vector $\xi_r$ is generated according to the probability distribution
\begin{equation}\label{xsidistr}
P(\xi_r) \propto 
\exp\{-\xi_r^{\dagger}Q_r^{\dagger}(V^{\prime})Q_r(V^{\prime})\xi_r\} \,.
\end{equation}
In practice, we start by generating a random Gaussian vector $R$ from which the
vectors
\begin{eqnarray}
 \Phi^{\prime} & = & Q^{\dagger}(V^{\prime})R \,, \label{Phip} \\
 X^{\prime} & = & [Q^{\dagger}(V^{\prime})Q(V^{\prime})]^{-1}\Phi^{\prime} \,.
 \label{Xp}
\end{eqnarray}
are constructed. The vector $\xi_r$ in \eq{xsidistr} is then given by
$\xi_r=A(V^{\prime})X^{\prime}$. The two terms in eq. (\ref{deltaS_r}) are
calculated with the formulae
\begin{eqnarray}
 \xi_r^{\dagger}Q^{\dagger}_r(V^{\prime})Q_r(V^{\prime})\xi_r & = & 
 \Phi^{\prime\dagger}X^{\prime} \,, \label{dSr_new} \\
 \xi_r^{\dagger}Q^{\dagger}_r(V)Q_r(V)\xi_r & = &
 X^{\prime\dagger}A(V^{\prime})A(V)^{-1}Q^{\dagger}(V)Q(V)
 A(V)^{-1}A(V^{\prime})X^{\prime} \,. \label{dSr_old}
\end{eqnarray}
All the vectors $\Phi^{\prime}$, $X^{\prime}$, $\xi_r$ and the traces in
eq. (\ref{seffD4}) are defined on the even sites of the lattice only.

The real parameters $\alpha_4$ and $\alpha_2$ can be optimized to maximize
$\tilde{P}_{\rm acc}$, eq. (\ref{tildepacc}).
We keep the choice $\alpha_4=-0.006$, $\alpha_2=-0.18$ from our previous work
\cite{Knechtli:2000ku} where the values were found to be optimal for an
action with smeared links obtained by three levels of ordinary APE blocking
\cite{Albanese:1987ds}. The computation of the matrix $A$ in eq. (\ref{amat})
requires the evaluation of the exponential function which we do by truncating
the exponential series after 15 terms.

In Monte Carlo simulations we separate the measurements of the observables
by $N_{\rm OR}$ overrelaxation and $N_{\rm HB}$ heatbath two--step updatings,
changing $t_{\rm OR}$ and $t_{\rm HB}$ links respectively. On $8^3\times24$
lattices we simulated the HYP action with gauge coupling $\beta=5.2$ and bare quark
mass $am=0.1$. An approximate matching of the lattice spacing 
($a\sim0.17\,{\rm fm}$) and of the Goldstone pion mass ($m_{\rm G}r_0=2.0$)
can be achieved by simulating the thin link action
with parameters $\beta=5.2$ and $am=0.06$. In an attempt to make a reliable
comparison of the computational cost of the two simulations, we give in the last
column of table \ref{t_cost} estimates for the number of $Q^{\dagger}Q$
multiplications required to get an independent configuration. These estimates are
obtained by considering the autocorrelation time $\tau_{\rm int}(p)$ for the
plaquette. By independent configurations we mean configurations obtained every
$2\tau_{\rm int}(p)$ updating steps.
\begin{table}[h]
\vspace{0.0cm}
 \caption{Comparison of the computational cost for creating an independent
configuration on a $10\,{\rm fm}^4$ lattice.
\label{t_cost}}
\vspace{0.2cm}
 \begin{tabular}{|c|c|c|c|c|c|} \hline
  Algorithm & $2\tau_{\rm int}$ & $Q^{\dagger}Q$/CG & 
  $Q^{\dagger}Q$/updating & upds./conf. &
  $Q^{\dagger}Q$/indep. conf.
  \\ \hline\hline
  thin HMC & 24(9)  & 135 & $67\times135\approx9000$ & 1 & $0.22(8)\times10^6$ 
  \\ \hline
  HYP      & 30(14) &  77 & 77+60+(33)=170 & 240 & $1.2(6)\times10^6$ 
  \\ \hline
 \end{tabular}
\vspace{0.0cm}
\end{table}

We found that measurements with the HYP action separated by
$N_{\rm OR}=160$ overrelaxation and $N_{\rm HB}=80$ heatbath updatings have
$\tau_{\rm int}(p)=15(7)$.
Each overrelaxation updating changes $t_{\rm OR}=128$ and each heatbath updating 
changes $t_{\rm HB}=200$ links with an acceptance of $\sim20\%$, corresponding
to an effective change of 15\% of all the links between measurements.
Each updating, with overrelaxation or heatbath, requires one inversion of
$Q^{\dagger}(V)Q(V)$, see eq. (\ref{Xp}), performed with one conjugate gradient
iteration (CG). Each CG iteration requires with these parameters on average 77
multiplications by $Q^{\dagger}(V)Q(V)$. In addition, each updating requires twice
the computation of the matrix $A(V)$, see eq. (\ref{dSr_old}), for a total of
60 multiplications by $Q^{\dagger}(V)Q(V)$. The recomputation of the HYP
links once 200 dynamical thin links are changed has a cost on this lattice volume
comparable to 33 multiplications by $Q^{\dagger}(V)Q(V)$. All together, we need 170
multiplications by $Q^{\dagger}(V)Q(V)$ for each updating.

We simulated the thin link action with HMC.
We found that measurements separated by one HMC trajectory of
unit time length have $\tau_{\rm int}(p)=12(5)$. One trajectory requires
about 67 CG iterations, each of which needs with the given parameters 
on average 135
multiplications by $Q^{\dagger}(U)Q(U)$.

In summary, an independent configuration for the HYP simulation costs
$1.2(6)\times10^6$ multiplications by $Q^{\dagger}(V)Q(V)$ and for the thin
link simulation $0.22(8)\times10^6$ multiplications by $Q^{\dagger}(U)Q(U)$.
This gives a factor 5(3) for the HYP computational cost over thin link HMC.
To achieve the same improvement of
flavor symmetry and scaling of the HYP action, the lattice spacing of thin
link action simulations should be reduced at least by a factor 2.
The overall gain of simulations with the HYP action is evident.

A drawback of our algorithm is that as the lattice volume increases the numbers
of the updated links $t_{\rm OR}$ and $t_{\rm HB}$ have to be kept unchanged,
consequently the numbers of updatings $N_{\rm OR}$ and $N_{\rm HB}$
have to be scaled with the volume to keep the autocorrelation times unchanged.
Since each updating requires the evaluation of the fermionic action this gives
a volume square dependence for the cost of the algorithm at fixed lattice spacing.
On the other hand as the continuum limit is approached the numbers of
links $t_{\rm OR}$ and $t_{\rm HB}$ which can be effectively updated scale: 
the physical volume of the updated region is constant.
Considering our simulation of the HYP action at $\beta=5.2$ and $am=0.1$,
we can update a physical volume
$0.027(1)\,{\rm fm}^4$ with one overrelaxation updating and 
$0.042(2)\,{\rm fm}^4$ with one heatbath updating with 20\% acceptance rate.
Simulating the HYP action at $\beta=5.0$ and $am=0.1$, with a
lattice spacing which is 30-40\% larger, we can update a physical volume
$0.034(6)\,{\rm fm}^4$ with one overrelaxation updating and 
$0.056(10)\,{\rm fm}^4$ with one heatbath updating with 20\% acceptance rate.

To improve the efficiency of the HYP update one can improve the 
stochastic estimator needed in the acceptance step, adding a $D^6$ term
in the definition of the matrix $A$ eq. (\ref{amat}). Furthermore, the pure gauge
action used in the heatbath and overrelaxation updatings can be changed to make
a better proposal for the acceptance step and the convergence of the
exponential series calculation for the matrix $A$ can be made faster. 
Some of these improvements are being discussed in a forthcoming publication
\cite{annanew}.

\section{Conclusions \label{s_concl}}

We presented a new simulation algorithm applicable to QCD actions where the
fermions are coupled via smeared links. The algorithm is based on the standard
overrelaxation and heatbath algorithms. We used it to simulate
four flavors of staggered fermions with HYP smeared links. 
The cost of the algorithm has a volume square dependence at fixed lattice spacing
due to the finite volume of the lattice region that can be updated,
but the physical volume of the updated region remains constant
as the continuum limit is approached. A comparison with the time cost of
standard staggered action simulations with Hybrid Monte Carlo shows that the
new algorithm is a factor 2--8 slower on lattices of physical volume of 
about $10\,{\rm fm}^4$ at $m_Gr_0=2.0$. The improvement in flavor symmetry
and scaling of the HYP action allows to gain a factor 2 in the lattice spacing $a$.
Given that the present cost of QCD simulations grows at least as $1/a^6$ there
is a consistent improvement with the new algorithm.

An interesting point is that the algorithm
could be used to simulate chiral fermionic actions like overlap or perfect actions.
The presence of smeared links is the key feature which makes the algorithm
efficient. 

{\bf Acknowledgement.} We are indebted to M. Hasenbusch for discussions that
led to the new algorithm. We benefited from many discussions with Prof.
T. DeGrand during the course of this work. We thank Prof. D. Toussaint for sharing 
with us the quenched Asqtad data of the MILC collaboration. 
The simulations were performed
in scalar and parallel mode on the beowulf cluster of the high energy theory
group and on the linux farm of the high energy experimental group at University
of Colorado. We would like to thank the system administrator, D. Johnson, for
his technical support. This work was supported by the U.S. Department
of Energy. Finally we thank the MILC collaboration for the use of their computer
code.

   \bibliography{lattice}        

\begin{thebibliography}{10}

\bibitem{Orginos:1998ue}
MILC, K. Orginos and D. Toussaint,
\newblock Phys. Rev. D59 (1999) 014501, hep-lat/9805009.

\bibitem{Orginos:1999cr}
MILC, K. Orginos, D. Toussaint and R.L. Sugar,
\newblock Phys. Rev. D60 (1999) 054503, hep-lat/9903032.

\bibitem{Knechtli:2000ku}
F. Knechtli and A. Hasenfratz,
\newblock Phys. Rev. D63 (2001) 114502, hep-lat/0012022.

\bibitem{DeGrand:1998mn}
T. DeGrand, A. Hasenfratz and T.G. Kovacs,
\newblock Nucl. Phys. B547 (1999) 259, hep-lat/9810061.

\bibitem{Bietenholz:2000iy}
W. Bietenholz,
\newblock (2000), hep-lat/0007017.

\bibitem{DeGrand:2000tf}
MILC, T. DeGrand,
\newblock Phys. Rev. D63 (2001) 034503, hep-lat/0007046.

\bibitem{DeGrand:2000gq}
T. DeGrand and A. Hasenfratz,
\newblock Phys. Rev. D64 (2001) 034512, hep-lat/0012021.

\bibitem{DeGrand:1998pr}
MILC, T. DeGrand,
\newblock Phys. Rev. D58 (1998) 094503, hep-lat/9802012.

\bibitem{Niedermayer:2000yx}
F. Niedermayer, P. Rufenacht and U. Wenger,
\newblock Nucl. Phys. B597 (2001) 413, hep-lat/0007007.

\bibitem{Hasenfratz:2001hr}
P. Hasenfratz, S. Hauswirth, K. Holland, T. Jorg and F. Niedermayer,
\newblock (2001), hep-lat/0109004.

\bibitem{DeGrand:1995ji}
T. DeGrand, A. Hasenfratz, P. Hasenfratz and F. Niedermayer,
\newblock Nucl. Phys. B454 (1995) 587, hep-lat/9506030.

\bibitem{Hasenfratz:2001tw}
A. Hasenfratz, R. Hoffmann and F. Knechtli,
\newblock (2001), hep-lat/0110168.

\bibitem{Gattringer:2001jf}
C. Gattringer, R. Hoffmann and S. Schaefer,
\newblock (2001), hep-lat/0112024, accepted for publication in Phys. Rev. D.

\bibitem{Karsch:2000ps}
F. Karsch, E. Laermann and A. Peikert,
\newblock Phys. Lett. B478 (2000) 447, hep-lat/0002003.

\bibitem{Bernard:2001av}
C.W. Bernard et~al.,
\newblock Phys. Rev. D64 (2001) 054506, hep-lat/0104002.

\bibitem{Hasenfratz:2001hp}
A. Hasenfratz and F. Knechtli,
\newblock Phys. Rev. D64 (2001) 034504, hep-lat/0103029.

\bibitem{annapert}
A. Hasenfratz,
\newblock in preparation .

\bibitem{Sommer:1994ce}
R. Sommer,
\newblock Nucl. Phys. B411 (1994) 839, hep-lat/9310022.

\bibitem{dougasqtad}
D. Toussaint,
\newblock private communication .

\bibitem{Gupta:1991mr}
R. Gupta, G. Guralnik, G.W. Kilcup and S.R. Sharpe,
\newblock Phys. Rev. D43 (1991) 2003.

\bibitem{Hasenbusch:1998yb}
M. Hasenbusch,
\newblock Phys. Rev. D59 (1999) 054505, hep-lat/9807031.

\bibitem{Adler:1981sn}
S.L. Adler,
\newblock Phys. Rev. D23 (1981) 2901.

\bibitem{Petronzio:1990vx}
R. Petronzio and E. Vicari,
\newblock Phys. Lett. B245 (1990) 581.

\bibitem{Cabibbo:1982zn}
N. Cabibbo and E. Marinari,
\newblock Phys. Lett. B119 (1982) 387.

\bibitem{Martin:1985yn}
O. Martin and S.W. Otto,
\newblock Phys. Rev. D31 (1985) 435.

\bibitem{Albanese:1987ds}
APE, M. Albanese et~al.,
\newblock Phys. Lett. B192 (1987) 163.

\bibitem{annanew}
A. Alexandru and A. Hasenfratz,
\newblock in preparation .

\end{thebibliography}
   \bibliographystyle{h-elsevier}   
\end{document}